\documentclass[12pt]{article}
\usepackage{latexsym}
\usepackage{epsfig,amssymb,euscript}
\usepackage{amsmath}
\numberwithin{equation}{section}  
\newsavebox{\ns}
\newsavebox{\dbrane}
\newsavebox{\dbshort}

\def\be{\begin{equation}}
\def\ee{\end{equation}}
\def\bea{\begin{eqnarray}}
\def\eea{\end{eqnarray}}

\newcommand{\nn}{\nonumber}

\def\Dslash{\,\,{\raise.15ex\hbox{/}\mkern-12mu D}}
\def\Dbarslash{\,\,{\raise.15ex\hbox{/}\mkern-12mu {\bar D}}}
\def\delslash{\,\,{\raise.15ex\hbox{/}\mkern-9mu \partial}}
\def\delbarslash{\,\,{\raise.15ex\hbox{/}\mkern-9mu {\bar\partial}}}
\def\pslash{\,\,{\raise.15ex\hbox{/}\mkern-9mu p}}
\def\calDslash{\,\,{\raise.15ex\hbox{/}\mkern-12mu {\cal D}}}

\newcommand\R{\mathbb{R}}
\newcommand\Z{\mathbb{Z}}

\newcommand\C{\mathbb{C}}
\newcommand\T{\mathbb{T}}
\newcommand\diff{\mathrm{d}}
\newcommand{\de}{\partial}
\newcommand{\Gcan}{G^{\mathrm{can}}}
\newcommand{\bcan}{b^{\mathrm{can}}}
\newcommand{\vol}{\mathrm{vol}}

\begin{document}
\begin{titlepage}
\begin{center}
\today
{\small\hfill hep-th/0503183}\\
{\small\hfill CERN-PH-TH/2005-047}\\
{\small\hfill HUTP-05/A0012}\\

\vskip 2.5 cm 
{\Large \bf The Geometric Dual of $a$--maximisation for }\\
\vskip 4mm
{\Large \bf Toric Sasaki--Einstein Manifolds}

\vskip 7mm
{Dario Martelli$^{1}$,~~ James Sparks$^{2,3}$~~and~~ Shing--Tung Yau$^{2}$}\\
\vskip 1 cm

1: Department of Physics, CERN Theory Division\\
1211 Geneva 23, Switzerland\\
\vskip 0.5cm
2: Department of Mathematics, Harvard University \\
One Oxford Street, Cambridge, MA 02138, U.S.A.\\
\vskip 0.5cm
3: Jefferson Physical Laboratory, Harvard University \\
Cambridge, MA 02138, U.S.A.\\

\vskip 5mm
{\tt dario.martelli@cern.ch  \ \  sparks@math.harvard.edu }\\
{\tt yau@math.harvard.edu}\\

\end{center}

\begin{abstract}
\noindent
We show that the Reeb vector, and hence in particular the volume, of a 
Sasaki--Einstein metric on the base of a toric Calabi--Yau cone
of complex dimension $n$ 
 may be computed by minimising a function $Z$ on $\R^n$
which depends only on the toric data that defines the singularity. 
In this way one can extract certain 
geometric information for a toric Sasaki--Einstein 
manifold without finding the metric explicitly.
For complex dimension $n=3$ the Reeb vector and the volume correspond to 
the R--symmetry and the $a$ central charge of the AdS/CFT dual superconformal 
field theory, respectively. 
We therefore interpret this extremal problem as the 
geometric dual of $a$--maximisation. 
We illustrate our results with some examples, including 
the $Y^{p,q}$ singularities and 
the complex cone over the second del Pezzo surface. 

\end{abstract}

\end{titlepage}
\pagestyle{plain}
\setcounter{page}{1}
\newcounter{bean}
\baselineskip18pt


\section{Introduction}

There has been considerable interest recently in Sasaki--Einstein 
geometry. Recall that a Sasaki--Einstein manifold $Y$ is a Riemannian 
manifold of dimension $(2n-1)$ whose metric cone
\be
\diff s^2 (C(Y)) = \diff r^2 + r^2 \diff s^2 (Y)\label{metcone}\ee
is Ricci--flat and K\"ahler. The recent interest has largely arisen due 
to a new construction of explicit inhomogeneous Sasaki--Einstein metrics 
in all 
dimensions \cite{paper1,paper2, paper3}. In particular in dimension $n=3$ there is an infinite family of cohomogeneity one 
five--metrics, denoted $Y^{p,q}$ where $q<p$ are positive integers 
\cite{paper2}. The  AdS/CFT correspondence \cite{Maldacena}
conjectures that for a Sasaki--Einstein five--manifold $Y$, type IIB string theory
on AdS$_5\times Y$ with $N$ units of self--dual
five--form flux is dual to a four--dimensional $\mathcal{N}=1$
superconformal field theory \cite{Kehagias,KW,acharya,MP}. This field theory 
may be thought of as arising from a stack of $N$ D3--branes
sitting at the apex $r=0$ of the corresponding Calabi--Yau cone (\ref{metcone}). 
Following the results of \cite{toricpaper},  for the case $Y=Y^{p,q}$ these field 
theories were constructed in \cite{quiverpaper} thus furnishing a countably infinite set of 
AdS/CFT duals where both sides of the duality are known explicitly.

Recall that all Sasaki--Einstein manifolds $Y$ have a canonically 
defined constant norm Killing vector field $K$, called the Reeb vector. 
In the case $n=3$ this is AdS/CFT dual to the R--symmetry of the 
dual superconformal field theory. The transverse geometry to the corresponding 
foliation of $Y$ is always K\"ahler--Einstein of positive curvature. 
In the case that the leaves of the foliation are all compact
one has a $U(1)$ action on $Y$. If this action is free the 
Sasaki--Einstein manifold is said to be regular, and 
is the total space of a $U(1)$ principle 
bundle over a positive curvature K\"ahler--Einstein manifold. 
More generally the $U(1)$ action is only locally free and one instead
has a $U(1)$ orbibundle over a positive curvature K\"ahler--Einstein 
orbifold. Such structures are referred to as quasi--regular. 
If the generic orbits of $K$ do not close there is only 
a transverse K\"ahler--Einstein structure and these are the irregular 
geometries.

In dimension five, regular Sasaki--Einstein metrics are 
classified completely  ~\cite{fried}. This follows since the smooth
four--dimensional K\"ahler--Einstein metrics with positive
curvature on the base have been classified by Tian and Yau
\cite{tian,tianyau}. These include the special cases $\mathbb{C}P^2$ and
$S^2\times S^2$, with corresponding Sasaki--Einstein manifolds
being the homogeneous manifolds $S^5$ (or $S^5/\Z_3$) and
$T^{1,1}$ (or $T^{1,1}/\Z_2$), respectively. 
For the remaining metrics the base is a del Pezzo surface
obtained by blowing up $\mathbb{C}P^2$ at $k$ generic points with $3\le
k\le8$ and, although proven to exist, the general metrics are not 
known in explicit form. In the last few years, starting with the work of
Boyer and Galicki \cite{boyerone},  quasi--regular Sasaki--Einstein 
metrics have been shown to exist on $\#l(S^2\times S^3)$ with $l=1,\dots, 9$. 
The irregular case is perhaps more interesting since so little is known about 
these geometries -- the $Y^{p,q}$ metrics \cite{paper2} and their 
higher dimensional 
generalisations \cite{paper3, general1, general2} are the very first examples. 
Indeed, these are counterexamples to the conjecture of 
Cheeger and Tian \cite{cheeger} that irregular Sasaki--Einstein 
manifolds do not 
exist.

For an irregular metric the closure of the orbits of $K$ is at least a two--torus, meaning that 
the metric must possess at least a $U(1)\times U(1)$ group of isometries.
In this paper we restrict our 
attention to toric Sasaki--Einstein manifolds. By definition this 
means that the isometry group contains at least an $n$--torus. 
There are good mathematical and physical reasons for imposing toricity. 
On the mathematical side, as we shall see, the subject of 
toric Sasakian manifolds is simple enough that one can prove many 
general results without too much effort. 
On the physical side, for $n=3$, a toric Sasaki--Einstein manifold is dual to a 
\emph{toric quiver gauge theory}.  These theories have a rich structure, but again are simple enough that 
one has considerable analytic control.  

Given a Sasaki--Einstein five--manifold $Y$, the problem of 
constructing the dual field theory is in general a difficult one. However, provided 
the isometry group of $Y$ is large enough one can typically 
make progress using a variety of physical and mathematical 
arguments. In particular, if $Y$ is toric in principle\footnote{In 
practice this algorithm requires a computer, and even then one is 
limited to relatively small -- in the sense of the toric 
diagram -- singularities.}
there is an algorithm which constructs the gauge theory from the 
toric data of the Calabi--Yau singularity \cite{hanany,hanany2}. Thus in this case 
both the geometry and the gauge theory are specified by a set of 
combinatoral data. On physical grounds, this theory is expected 
to flow at low energies to a superconformal fixed point, and in 
particular the global symmetry group of this theory 
contains a canonical ``$U(1)_R$'' factor, which is the R--symmetry.
If this symmetry is correctly identified, many properties of the gauge theory
may be determined. A general procedure that determines this symmetry is 
 $a$--maximisation \cite{IW}.
Roughly, one can define a function $a$ on an appropriate space of admissable 
R--symmetries which depends only on the combinatorial data that 
specifies the quiver gauge theory, and thus 
in principle only on the toric data of the singularity. The local maximum of this 
function precisely determines the R--symmetry of the theory at its 
superconformal point. From the R--charges one can then use the 
AdS/CFT correspondence to compute the volume of the dual 
Sasaki--Einstein manifold, as well as the volumes of certain supersymmetric 
3--dimensional submanifolds.  
Remarkable agreement was found for these
two computations in the case of the $Y^{2,1}$ metric \cite{toricpaper}, 
and the $a$--maximisation 
calculation \cite{BBC} for the quiver gauge theory corresponding to the first del Pezzo surface  \cite{hanany}. 
The field theories for the remaining $Y^{p,q}$ family 
were 
constructed in \cite{quiverpaper} and again 
perfect agreement was found for the two computations.

To summarise, $a$--maximisation and the AdS/CFT correspondence imply
that the volumes of toric Sasaki--Einstein manifolds, as well as certain submanifolds, 
should somehow be extractable from the toric data of the 
Calabi--Yau singularity in a relatively simple 
manner, without actually finding the metric. 
In both the regular and quasi--regular cases this follows from the 
fact that, in these cases, one can view the Sasaki--Einstein 
manifold as a $U(1)$ (orbi)--bundle
over a K\"ahler--Einstein manifold (respectively orbifold), 
where the $U(1)$ is
generated by the Reeb vector.  The problem of computing 
the volume, as well as the volumes of certain supersymmetric submanifolds, 
is then reduced to that
of computing the volumes of the K\"ahler--Einstein base and its 
divisors, respectively,
which is a purely topological question, see \emph{e.g.} \cite{besse}. 
These are then clearly \emph{rational} multiples of the volumes
of the round five--sphere and three--sphere, respectively. 
However, in some sense the generic case is the irregular 
case and here one cannot reduce the computation to that of computing 
topological invariants. In 
this paper we show that 
one can determine the Reeb vector of any toric Sasaki--Einstein 
manifold in a simple way, without finding the metric, and from 
this one can compute the volumes referred to above. We therefore interpret 
this as being a geometric ``dual'' to $a$--maximisation.

\section{Toric Sasakian Geometry}

In this section we describe 
the K\"ahler geometry of toric varieties, focusing 
on the special case of a K\"ahler cone. The general
formalism is due to Guillemin \cite{Guillemin} and Abreu \cite{abreu1} 
and has been used recently 
in Donaldson's work \cite{Donaldson1, Donaldson2} on constant scalar curvature metrics. 
Here we focus on the case where the K\"ahler toric variety is a cone over 
a real manifold, which by definition is a Sasakian manifold. 
The torus action fibres this K\"ahler cone over a rational polyhedral cone 
$\mathcal{C}\subset\R^n$ via the moment map.
Any toric 
K\"ahler metric may be written in terms of a symplectic potential, which 
is the Legendre transform of the K\"ahler potential, and in the 
special case of a cone we show that the moduli space of such symplectic 
potentials, for fixed toric variety, splits as 
\be
\mathcal{S} = \mathcal{C}^*_0\times \mathcal{H}(1)\label{bob}\ee
where $\mathcal{C}^*_0$, the space of Reeb vectors, is the interior of the
dual cone to $\mathcal{C}$ and 
$\mathcal{H}(1)$ is the space of smooth homogeneous degree one 
functions on $\mathcal{C}$ (subject to a convexity condition). 
We also write down a Monge--Amp\`ere equation in this formalism 
which imposes that the Sasakian metric is also Einstein. Regularity 
of a solution to this equation then imposes a condition on the 
Reeb vector $K$. 

\subsection*{Sasakian Geometry} 

Let $(X,\omega)$ be a K\"ahler cone of complex dimension $n$. 
This means that $X=C(Y)\cong \R^+\times Y$ 
has metric
\be
\diff s^2(X) = \diff r^2 + r^2 \diff s^2(Y)~.\ee
We take $r>0$ so that $X$ is a smooth manifold which is incomplete at 
$r=0$. The condition that this metric be K\"ahler is then equivalent to 
$Y=X\mid_{r=1}$ being Sasakian -- in fact this is probably the most useful 
definition of Sasakian. We then have
\be
\mathcal{L}_{r\de/\de r} \omega = 2\omega\ee
which says that the K\"ahler form $\omega$ is homogeneous degree 2 under the 
Euler vector $r\de/\de r$. It follows that $\omega$ is exact:
\be
\omega = -\frac{1}{2}\diff (r^2\eta)\ee
where $\eta$ may be considered as a global one--form on $Y=X\mid_{r=1}$. 

From this definition it is straightforward to show that the Reeb vector field
\be
K\equiv \mathcal{I}\left(r\frac{\de}{\de r}\right)\label{Killbill}\ee
is a Killing vector field, where $\mathcal{I}$ denotes the 
complex structure on $X$.  
$K$ is dual to the one--form $r^2\eta$, as follows simply 
from the above definitions. 
Thus equivalently we have
\be
\eta = \mathcal{I}\left(\frac{\diff r}{r}\right)~.\ee
It terms of the $\de$ operator on $X$ we thus have
\be
\eta = i(\de-\bar{\de})\log r\ee
so that 
\be
\diff \eta = -2i \de\bar{\de}\log r~.\ee
Moreover one now computes that the K\"ahler form is simply
\be
\omega = \frac{1}{2}i\de\bar{\de}r^2\ee
and thus we see that $F\equiv r^2/4$ is 
a K\"ahler potential. 

\subsection*{Symplectic point of view}

We now impose in addition that $(X,\omega)$ is toric. This means that
the real torus $\T^n$ acts effectively on $X$, preserving the K\"ahler form, 
which we regard as a symplectic form. Moreover one also requires that the 
torus action is integrable, meaning 
that one can introduce a moment map $\mu:X\rightarrow \R^n$. 
The moment map allows one to introduce 
symplectic coordinates on $\R^n$
\be
y_i = -\frac{1}{2}<r^2\eta,\frac{\de}{\de\phi_i}>\ee
where $\partial/\partial\phi_i$ generate the 
$\T^n$ action. Thus $\phi_i$ are angular coordinates along 
the orbits of the torus action, with
$\phi_i\sim\phi_i+2\pi$. We may then use $(y,\phi)$ as 
symplectic coordinates on $X$. Let us also 
assume\footnote{The symplectic toric cones that are \emph{not}
of Reeb type are rather uninteresting: they are either cones over
$S^2\times S^1$, cones over 
principle $\T^3$ bundles over $S^2$, or cones over products $\T^m\times S^{m+2j-1}$,
$m>1, j\geq 0$ \cite{L}.} that $X$ is of
Reeb type. This means that there is some $\zeta$ 
such that $-<r^2\eta,\zeta>$ is a strictly positive
function on $X$. The moment map then exhibits 
the K\"ahler cone as a Lagrangian torus fibration 
over a strictly convex rational polyhedral cone 
$\mathcal{C}\subset \R^n$ by forgetting the angular coordinates
$\phi_i$ \cite{FT}. This image is a subset of $\R^n$ of the form
\be\label{conehead}
\mathcal{C} = \{y\in\R^n\mid l_a(y)\geq 0, a=1,\ldots,d\}\ee
where we have introduced the linear function
\be
l_a(y)=(y,v_a)\ee
with Euclidean metric $(\cdot,\cdot)$, 
and $v_a$ are the inward pointing normal vectors to the $d$ facets of the 
polyhedral cone. These normals are rational and hence one can normalise 
them to be primitive\footnote{A vector $v\in\Z^n$ is primitive 
if it cannot be written as $mv^{\prime}$ with 
$v^{\prime}\in\Z^n$ and $\Z\ni m>1$.} 
elements of $\Z^n$. We also assume this set of vectors is 
minimal in the sense that removing any vector $v_a$ in 
the definition (\ref{conehead}) changes $\mathcal{C}$.
The condition that $\mathcal{C}$ be strictly convex is simply 
the condition that it is a cone over a convex polytope. There is an 
additional condition on the $\{v_a\}$ for $Y$ a smooth manifold, and the 
cone is then said to be \emph{good} \cite{L}. This may be defined as follows. 
Each face $\mathcal{F}\subset\mathcal{C}$ may be realised uniquely as the 
intersection of some number of facets $\{l_a(y)=0\}$. 
Denote by $v_{a_1},\ldots,v_{a_N}$ the corresponding 
collection of normal vectors in $\{v_a\}$, where $N$ is the 
codimension of $\mathcal{F}$ -- thus $\{a_1,\ldots,a_N\}$ is a 
subset of $\{1,\ldots,d\}$. Then the cone is good if and only if 
\be\label{good}
\left\{\sum_{A=1}^N {\nu}_A v_{a_A}\mid {\nu}_A\in \R\right\}\cap\Z^n 
= \left\{\sum_{A=1}^N {\nu}_A v_{a_A}\mid {\nu}_A\in \Z\right\}\ee
for all faces $\mathcal{F}$.

The torus fibration 
is non--degenerate over the interior $\mathcal{C}_0$ 
of $\mathcal{C}$. Thus the $\T^n$ action is free 
on the corresponding subset $X_0 = \mu^{-1}(\mathcal{C}_0)$ of $X$.
The boundary $\partial\mathcal{C}$ of the polyhedral cone then effectively
describes $X$ as a compactification of 
$\mathcal{C}_0 \times \T^n$. Specifically, 
the normal vector $v_a\in\Z^n$ to a facet $\{l_a(y)=0\}$ determines 
a one--cycle in $\T^n$ and this cycle collapses over the facet. 
Thus each facet corresponds to a toric symplectic 
subspace of $X$ of real codimension two. Similarly lower--dimensional 
faces of the cone correspond to higher codimension toric symplectic 
subspaces. The condition that the cone is good then amounts to requiring 
that this compactification gives a cone over a smooth manifold $Y$.

The symplectic (K\"ahler) form is 
\be
\omega = \diff y_i\wedge \diff\phi_i\ee
where here and henceforth 
we adopt the Einstein summation convention for the indices 
$\{i,j,k,\ldots\}$. As described in \cite{abreu1}, any $\T^n$--invariant 
K\"ahler metric on $X$ is then of the form
\be\label{symplecticmetric}
\diff s^2 =  G_{ij} \diff y_i\diff y_j 
+ G^{ij}\diff \phi_i\diff \phi_j\ee
where $G^{ij}$ is the inverse matrix to 
$G_{ij}=G_{ij}(y)$. The almost complex structure is then clearly
\be\label{complex}
\mathcal{I} = \left[\begin{array}{cc}0 & -G^{ij}\\ G_{ij} & 0\end{array}\right]\ee
in the basis $(y,\phi)$ 
and it is straightforward to verify that integrability of $\mathcal{I}$
requires $G_{ij,k}=G_{ik,j}$ and hence 
\be
G_{ij} = G_{,ij}\equiv \frac{\partial^2G}{\partial y_i \partial y_j}\ee
for some strictly convex function $G = G(y)$. We refer to 
$G$ as the symplectic potential for the K\"ahler metric. 
It should be clear that the metric (\ref{symplecticmetric}) is a cone 
if and only if the matrix $G_{ij}(y)$ is homogeneous degree $-1$ in 
$y$.

\subsection*{Complex point of view}

The introduction of the symplectic potential $G(y)$ above may 
seem slightly mysterious, but in fact it is related to the 
more usual K\"ahler potential by Legendre transform. In fact the 
two viewpoints may be neatly summarised as follows. 
In the complex viewpoint one keeps the complex structure of $X$ fixed
and considers the K\"ahler form, and hence K\"ahler potential, to vary, whereas
in the symplectic viewpoint one keeps the symplectic form fixed 
and varies the complex structure (\ref{complex}). Usually this 
latter approach is not particularly useful in K\"ahler geometry. 
However in toric K\"ahler geometry 
this formalism has already been used with great success, for example 
in Donaldson's work \cite{Donaldson1, Donaldson2} on constant scalar curvature metrics.
This will also be the case for toric Sasakian metrics.

In the complex point of view one regards $X$ as a complex algebraic variety
coming equipped with a biholomorphic action of the complex torus
$\T^n_C = (\mathbb{C}^*)^n$ which has a dense open orbit $X_0$ which 
we identify  with $X_0$ above. We introduce standard complex 
coordinates $w_i$ on $\C\setminus \{0\}$. The real torus $\T^n\subset\T^n_C$
then acts by translation in the imaginary direction for the 
log complex coordinates $z_i = \log w_i = x_i + i\phi_i$. The 
K\"ahler form $\omega$ may then be written as
\be
\omega = 2i\partial\bar{\partial}F\ee
where $F=F(x)$ is the K\"ahler potential. Here we have again assumed that
the metric is invariant under the $\T^n$ symmetry. We also note 
that $F(x)$ is a strictly convex function of the 
variables $x$. In these coordinates the 
metric is
\be
\diff s^2 = F_{ij}\diff x_i\diff x_j + F_{ij}\diff \phi_i 
\diff \phi_j\ee
where 
\be
F_{ij}=\frac{\partial^2F}{\partial x_i \partial x_j}~.\ee
It follows that 
\be
F_{ij}(x) = G^{ij}(y=\partial F/\partial x)\ee
and the moment map is then clearly
\be
\mu = y = \frac{\partial F}{\partial x}\ee
by definition. 
It hence follows that the symplectic and K\"ahler potentials are related by 
Legendre transform
\be
F(x) = \left(y_i \frac{\partial G}{\partial y_i}-G\right)(y=\partial F/\partial x)~.\ee

\subsection*{Delzant construction and the canonical metric}

Given a good strictly convex rational 
polyhedral cone $\mathcal{C}\subset\R^n$ one can recover the 
original cone $X$, together with its symplectic structure, 
via symplectic reduction of $\C^d$. This follows from a generalisation \cite{L} of 
Delzant's theorem \cite{D}. In fact $X$ inherits 
a natural K\"ahler metric from K\"ahler reduction of the canonical metric 
on $\C^d$. The explicit formula for the symplectic potential of this metric
for compact K\"ahler toric varieties 
was first given in a beautiful paper of Guillemin \cite{Guillemin}. 
The case of singular varieties was studied recently in \cite{Lermansingular}.

Denote by $\Lambda \subset \Z^n$ the span
of the normals $\{v_a\}$ over $\Z$. This is a lattice of maximal rank. 
Consider the linear map 
\begin{eqnarray}\label{linear}
A: &&\R^d \rightarrow \R^n \nn \\
& & \ e_a\mapsto v_a\end{eqnarray}
which maps each standard orthonormal basis 
vector $e_a$ of $\R^d$ to the primitive normal vector $v_a$. 
This induces 
a map of tori
\be
\T^d\cong \R^d/2\pi\Z^d \rightarrow \R^n/2\pi\Lambda~.\ee
In general the 
kernel is $\mathcal{A}\cong \T^{d-n}\times \Gamma$ where $\Gamma$ is a finite 
abelian group. Then $X$ is given by the symplectic quotient 
\be
X = \C^d//\mathcal{A}~.\ee
One can describe this more explicitly as follows. One computes 
a primitive basis for the kernel of $A$ over $\Z$ by finding all 
solutions to 
\be
\sum_a Q_{I}^a  v_a = 0\ee
for $Q_{I}^a\in \Z$, and such that for each $I$ the $Q_I^a$ have no 
common factor. The number of solutions, indexed by $I$, is 
$d-n$ since $A$ is surjective -- this latter fact 
follows since $\mathcal{C}$ is strictly convex. Then one has 
\be\label{reduction}
X= \mathcal{K}/\T^{d-n}\times
\Gamma
\equiv \C^d//\mathcal{A}\ee
with
\be
\mathcal{K} \equiv \left\{(Z_1,\ldots,Z_d)\in\C^d\mid\sum_{a}Q_{I}^a |Z_a|^2 = 0\right\}\subset \C^d\ee
where $Z_a$ denote complex coordinates on $\C^d$ and the charge matrix 
$Q_{I}^a$ specifies the torus $\T^{d-n}\subset \T^d$. The quotient 
group $\T^d/\mathcal{A}\cong\T^n$ then acts symplectically on $X$ and by 
construction the image of the induced 
moment map $\mu:X\rightarrow\R^n$ is the polyhedral cone $\mathcal{C}$ 
that one began with. This is proven in \cite{L}.

Now $X$ inherits a K\"ahler metric from the flat metric on $\C^d$ via
the reduction (\ref{reduction}). Moreover from the latter equation we  
see that this induced metric is 
clearly invariant under homothetic rescaling of the $\{Z_a\}$ and thus 
this metric will be a conical metric on $X$.
There is an elegant expression for this metric, which in terms of the 
symplectic potential is given by \cite{Guillemin}
\be
\Gcan (y) = \frac{1}{2}\sum_a l_a(y)\log l_a(y)~.\ee
We also note the following formulae:
\be
\frac{\de\Gcan}{\de y_i} = \frac{1}{2}\sum_a [1+\log l_a(y)]v^a_i\ee
\be
\Gcan_{ij} = \frac{1}{2}\sum_a v^a_i v^a_j \frac{1}{l_a(y)}~.\ee
In particular note that $\Gcan_{ij}$ is homogeneous degree $-1$ which implies 
that the corresponding K\"ahler metric (\ref{symplecticmetric}) 
is a cone. Also notice that $\Gcan_{ij}$ has 
simple poles at each of the $d$ facets $l_a(y)=0$. This singular behaviour is
required precisely so that the metric on $\mathcal{C}_0\times \T^n$ 
compactifies to a smooth\footnote{When making such statements 
we always regard $X$ as having its apex deleted.} metric on $X$. 
As we shall see when we consider the Einstein condition 
for $G(y)$, the metric $\Gcan_{ij}(y)$ is never Ricci--flat for 
$d>n$. The case $d=n$ is the case that $X$ is locally $\C^n$.

\subsection*{The Reeb vector and moduli space of symplectic potentials}

Recall that on any K\"ahler cone $(X,\omega)$ there is a canonically 
defined Killing vector 
field $K$ defined  by (\ref{Killbill}). 
In particular $K$ has norm one at $Y=\{r=1\}$ and thus the orbits of $K$ on 
$Y$ define a foliation of $Y$. We refer to such a Sasakian structure 
as quasi--regular or irregular, depending on whether the generic 
orbits close or not, respectively. In the irregular case 
note that the isometry group is 
at least $\T^m$, $m\geq 2$, with
the orbits of the Killing vector filling out a dense subset of the
orbits of the torus action. Indeed, the isometry group of a compact 
Riemannian manifold is always a compact Lie
group.  Hence the orbits of a Killing vector field define a one--parameter
subgroup, the closure of which will always be an abelian subgroup 
and thus a torus. The dimension of the closure of the orbits, $m$, 
is called the rank. 

It is also straightforward to show that the Reeb vector always lies 
in the centre of the Lie algebra of the autmorphism group of $Y$ -- 
that is, the group of diffeomorphisms that preserve the 
Sasakian structure. 
To see this, suppose that 
the vector field $V$ generates a symmetry of the K\"ahler cone. 
This means that $V$ commutes with the Euler vector $r\de/\de r$ and 
satisfies
\be
\mathcal{L}_V \omega = 0, \quad \mathcal{L}_V \mathcal{I} = 0\ee
where $\mathcal{L}$ denotes the Lie derivative. 
In particular $V$ is an isometry of the metric\footnote{The converse 
need not be true. The isometry group of the round $S^5$ is $SO(6)$ but the 
group which preserves a chosen complex structure is $U(3)$.}. 
We now compute
\be
[V,K] = \mathcal{L}_V K = \mathcal{L}_V \left[\mathcal{I}\left(r\frac{\de}{\de r}\right)\right] = 0~.\ee
Hence $K$ commutes with $V$ for all $V$ and so $K$ lies in the 
centre of the automorphism group.  

For a toric Sasakian manifold we may write
\be
K = b_i \frac{\de}{\de\phi_i}
\label{defreeb}\ee
and regard $K$ as the vector $b\in\R^n$. Using 
\be
r\frac{\de}{\de r} = 2 y_i\frac{\de}{\de y_i}\ee
one easily computes that, for a given toric Sasakian manifold with 
symplectic potential $G$, we have
\be
b_i = 2G_{ij} y_j~.\ee
It is straightforward to check that $b$ is indeed a constant 
vector. For,
\be
\frac{\de}{\de y_k}b_i = 2y_j G_{ij,k} + 2G_{ik} = 
2\left(y_j\frac{\de}{\de y_j}\right)G_{ik} + 2G_{ik}=
0\ee
where we have used Euler's theorem and the fact that $G_{ik}$ is homogeneous 
degree $-1$. For the canonical metric one easily computes
\be
\bcan = \sum_a v^a~.\ee

Suppose now that two different symplectic potentials 
$G$, $G^{\prime}$ have the same Reeb vector $b\in\R^n$. 
Defining $g=G^{\prime}-G$ we have
\be\label{dude}
\left(y_j\frac{\de}{\de y_j}\right) \frac{\de}{\de y_i} g = 0\ee
so that $g_{,i}$ 
is homogeneous degree 0 for each $i$. 
It follows that $g\in \mathcal{H}(1)$ is homogeneous degree $1$, 
up to a constant. 
To see this, note that (\ref{dude}) implies
\be
\frac{\de}{\de y_i}\left[\left(y_j\frac{\de}{\de y_j}\right) g - g\right] = 0\ee
and hence
\be
y_j \frac{\de}{\de y_j} g = g + t~.\ee
where $t$ is a constant. The constant degree of freedom in $G$ is 
clearly irrelevant. Indeed note that $G^{\prime}_{ij}=G_{ij}$ if and only if 
\be
g=\lambda_iy_i+t\ee
where $\lambda_i$, $t$ are constants. Thus the symplectic
potential should be thought of as being defined up to a linear function. 

Conversely, if $g = (G^{\prime} -G)\in\mathcal{H}(1)$ then 
the two symplectic potentials $G^{\prime}$ and $G$ 
define the same Reeb vector and 
indeed their Hessians are homogeneous degree $-1$. 

Let us now define 
\be
G_b(y) = \frac{1}{2}l_b(y)\log l_b(y) - \frac{1}{2}l_{\infty}(y)
\log l_{\infty}(y)~.\ee
where 
\be
l_b(y) = (b,y)\ee
and 
\be
l_{\infty}(y)= (\bcan,y) = \sum_a (v_a,y)~.\ee
Provided the plane $l_b(y) = \nu>0$ intersects 
the polyhedral cone $\mathcal{C}$ to form a finite polytope, this function is 
a smooth function on $\mathcal{C}$. In fact this condition is that
\be
(b, u_{\alpha}) > 0\ee
where the $u_{\alpha}\in\Z^n$ are the generating 
edges of the cone $\mathcal{C}$. Indeed note that we may write
\be
\mathcal{C} = \left\{\sum_{\alpha} \lambda_{\alpha}u_{\alpha}\in\R^n\mid \lambda_{\alpha}\geq 0\right\}~.\ee
This identifies $\mathcal{C}^*=\{b\in\R^n\mid (b,u_{\alpha})\geq 0\}$ as 
the dual cone to $\mathcal{C}$, which is also a convex rational polyhedral 
cone by Farkas' Theorem.
Moreover, 
\be
2y_j \frac{\de}{\de y_j} \frac{\de}{\de y_i} G_b = b_i - \bcan_i\ee
and we may quite generally write any symplectic potential as 
\be
G = \Gcan + G_b + g\ee
where the Reeb vector for this potential is $b$, and $g$ is a 
homogeneous degree one function. Since $\Gcan$ already has the 
correct singular behaviour at the facets for the metric to 
compactify to a smooth metric on $X$, we simply require that 
$g$ be smooth and $b\in\mathcal{C}^*_0$ in order that this is also true for 
$G$. One also requires that $G$ be strictly convex in order that
the metric is positive definite.

We may  summarise our results thus far as follows: 
\begin{quote}
   \textsl{The moduli space, $\mathcal{S}$, 
of symplectic potentials corresponding to smooth 
Sasakian metrics 
on some fixed toric Sasakian manifold $Y$ can be naturally written as 
\be
\mathcal{S}=\mathcal{C}^*_0 \times \mathcal{H}(1)\ee
where $b\in \mathcal{C}^*_0\subset \R^n$ labels 
the Reeb vector for the Sasakian structure, and 
$g\in \mathcal{H}(1)$ is a 
smooth homogeneous degree one function on $\mathcal{C}$, such 
that $G$ is strictly convex.} 
\end{quote}

\subsection*{The Monge--Amp\`ere equation}

Let $F(x)$ denote the K\"ahler potential for a smooth metric 
on $X$, where recall that $x_i$ are the real parts of complex 
coordinates on $X$. As is well known, 
the Ricci--form corresponding to $F(x)$ is given by
\be
\rho = -i\partial\bar{\partial}\log\det (F_{ij})~.\ee
Thus Ricci--flatness $\rho=0$ gives
\be
\log \det (F_{ij}) = -2\gamma_i x_i + c
\label{logdetF}\ee
where $\gamma_i$ and $c$ are constants, and we have noted that 
any $\T^n$--invariant pluri--harmonic function is necessarily 
of the form of the right hand side. We may now take the 
Legendre transform of this equation to obtain
\be\label{MA}
\det (G_{ij}) = \exp\left(2\gamma_i\frac{\de G}{\de y_i} -c\right)~.\ee
We will refer to this as the Monge--Amp\`ere equation 
in symplectic coordinates.

Up until this point we have not imposed any Calabi--Yau condition 
on $X$. In particular if $c_1(X)$ is non--zero one certainly cannot 
find a  Ricci--flat metric. We thus henceforth 
take $X$ to be a toric Gorenstein singularity. This means that, 
by an appropriate $SL(n;\mathbb{Z})$ trasformation, one can take the 
normal vectors for the polyhedral cone to be 
\be
v_a=(1,w_a)\ee 
for all $a$, where $w_a\in\Z^{n-1}$. In particular note this this means 
that the charge vectors $Q_I^a$ satisfy
\be
\sum_a Q_I^a =0\ee
for each $I$ which in turn implies that $c_1(X)=0$. 
The plot of the vectors $w_a$ in $\Z^{n-1}$ is 
usually called the toric diagram in the physics literature, at 
least in the most physically relevant case of $n=3$. 

Note that (\ref{MA}) implies that
\be
-n=(b,\gamma)~.\label{harry}
\ee
This follows by taking the derivative of (\ref{MA}) along the Euler 
vector and the fact that the left hand side is homogeneous degree $-n$. 
One now easily computes the right hand side of the 
Monge--Amp\`ere equation. Up to a 
normalisation factor we have
\be\label{MA2}
\det (G_{ij}) = \prod_a \left[\frac{l_a(y)}{l_{\infty}(y)}\right]^{(v^a, 
\gamma)} 
[l_b(y)]^{-n}\exp\left(2\gamma_i \frac{\de g}{\de y_i}\right)~.\ee
Note that, since $g\in\mathcal{H}(1)$, the exponential is homogeneous 
degree 0, and hence the right hand side is indeed homogeneous 
degree $-n$. However, in order that $\det (G_{ij})$ has the correct singularity 
structure so that the corresponding K\"ahler metric 
is smooth, it must be of the form \cite{abreu1, Donaldson2}
\be
\det (G_{ij})= f(y)\prod_a [l_a(y)]^{-1}\ee
where $f(y)$ is everywhere smooth on $\mathcal{C}$ minus its apex.
Thus we see that
\be
(v_a,\gamma) = -1\ee
for all $a$. Clearly this is a very strong constraint and this is 
essentially where one sees $c_1(X)=0$. For, if $v_a=(1,w_a)$ then this 
is solved by taking
\be
\gamma = (-1,0,\ldots,0)~.\ee
In particular from (\ref{harry}) we obtain 
\be
b_1=n~.\ee

We conclude this subsection by deriving an expression for the 
holomorphic $(n,0)$--form $\Omega$ of the Ricci--flat metric on the 
Calabi--Yau cone.
In complex coordinates, the $(n,0)$--form may be written in the 
canonical form
\be
\Omega = e^{i\alpha}(\det F_{ij})^{1/2} \diff z_1 \wedge \dots \wedge \diff z_n~
\ee
where $\alpha$ is a phase which is fixed by requiring $\diff \Omega =0$.  
Using equation
 (\ref{logdetF})  we obtain the following expression:
\bea
\Omega & = & e^{x_1+i\phi_1} (\diff x_1+ i \diff \phi_1) \wedge \dots \wedge (\diff x_n +i \diff\phi_n)~.
\eea
%
Here we've set $c=0$. Now, using (\ref{defreeb}),  it is straightforward to derive the following:
\bea
{\cal L}_{K} \Omega & = & i\, n\, \Omega \label{omegachargen}\\
{\cal L}_ {\frac{\de} { \de \phi_i}}   \Omega & = & 0 \qquad \qquad i=2,\dots, n  ~.\label{omeganeutral}
\eea

\subsection*{The characteristic hyperplane and polytope}

Let us fix a toric Gorenstein singularity with polyhedral cone
$\mathcal{C}\subset\R^n$ and let 
$G$ be a symplectic potential with Reeb vector 
$b\in\mathcal{C}^*_0$. 
The Reeb vector has norm one at $Y=\{r=1\}$, which reads
\be
1= b_ib_jG^{ij} = 2b_iG_{jk}y_kG^{ij} = 2(b,y)~.\ee
Thus the base of the cone $Y$ at $r=1$ defines a hyperplane
\be
\left\{y\in\R^n\mid (b,y)=\tfrac{1}{2}\right\}\ee
with outward unit normal vector $b/|b|$. 
We call this the characteristic 
hyperplane for the Sasakian manifold \cite{boyertoric}. 
Since $b\in\mathcal{C}^*_0$ this 
hyperplane intersects $\mathcal{C}$ to form a finite polytope 
$\Delta=\Delta_b$. We denote
\be
H = \left\{y\in\R^n\mid (b,y)=\tfrac{1}{2}\right\}\cap\mathcal{C}~.\ee
Note 
that the Sasakian manifold $Y$ is a $\T^n$ fibration over $H$. 
Notice also that 
the Sasakian structure is quasi--regular if and only if 
$b\in\mathbb{Q}^n$ is a rational point. 
One can interpret $H$ as a Delzant--Lerman--Tolman polytope \cite{LT}
if and only if the structure is quasi--regular and thus this 
polytope is rational.

Let us denote 
\be
X_1 = X\mid_{r\leq1}\ee
so that $X_1$ is a finite cone over the base $Y$. 
Correspondingly the image
\be
\mu(X_1)=\Delta=\Delta_b\ee
under the moment map is the finite polytope $\Delta$, which depends 
on the Reeb vector $b$. The volume of $X_1$ is
\be
\vol(X_1)=\int_0^1 \diff r\ r^{2n-1} \vol(Y) = \frac{1}{2n}\vol(Y)~.\ee
On the other hand, since $X$ is K\"ahler the volume form on 
$X$ is simply $\omega^n/n!$. Integrating this over 
$X_1$ one obtains
\be
\int_{\mu^{-1}(\Delta)} \frac{1}{n!}\omega^n = \int_{\mu^{-1}(\Delta)} 
\diff y_1\ldots \diff y_n \diff \phi_1\ldots\diff\phi_n 
= (2\pi)^n \vol(\Delta)\ee
where $\vol(\Delta)$ is simply the Euclidean volume 
of the polytope $\Delta$. We thus have the simple result
\be\label{vol}
\vol(Y) = 2n (2\pi)^n\vol(\Delta)~.\ee
Note that this depends only on $b$, for fixed toric singularity, and 
not on the homogeneous degree one function $g$.

Let us now consider toric divisors in $X$. These are just the 
inverse images of the facets of $\mathcal{C}$. To see this, note 
that each facet is the reduction of 
$\{Z_a=0\}\subset\C^d$ in Delzant's construction, which
clearly descends to a complex subspace of $X$. 
Thus each facet is the image under $\mu$ of 
a toric divisor $D_a$ in $X$. In particular the latter is 
calibrated by the form 
$\omega^{n-1}/(n-1)!$. A similar reasoning to the above then 
gives
\be\label{faces}
\vol(\Sigma_a) = (2n-2)(2\pi)^{n-1}\frac{1}{|v_a|}\vol(\mathcal{F}_a)\ee
where $\mathcal{F}_a=\{l_a(y)=0\}\cap \{r\leq 1\}$, 
$v_a$ is the primitive normal vector,  
and $\Sigma_a = \mu^{-1}(\mathcal{F}_a)\mid_{r=1}$ is the corresponding 
$(2n-3)$--submanifold of $Y$. Thus $D_a=C(\Sigma_a)$.

\begin{quote} 
\textsl{To summarise, the 
volumes $\vol(Y)$ and $\vol(\Sigma_a)$ 
depend only on the Reeb vector $b\in\mathcal{C}^*_0$ and not on the 
homogeneous degree one function $g$.}
\end{quote}
This will 
be especially important when we consider Sasaki--Einstein metrics. 
In this case it is a very difficult problem to find $b$ and the function 
$g$ which satisfy the Monge--Amp\`ere equation (\ref{MA2}). However, 
as we shall demonstrate shortly, these two components essentially 
decouple from each other, and one can determine $b$ 
for the Sasaki--Einstein metric independently of determining the 
function $g$.

\subsection*{A formula for the integrated Ricci scalar}

According to \cite{abreu2} we have the following 
formula for the Ricci scalar\footnote{We use 
a subscript $X$ to distinguish this from the 
Ricci scalar of the Sasakian metric which will appear 
presently. Obviously the two are closely 
related.} $R_X$ of a toric K\"ahler metric on 
$X$ in terms 
of the symplectic potential $G$:
\be
R_X = - G^{ij}_{\ \ ij} \equiv -G^{ij}_{\ \ ,ij}~.\ee
Let us now integrate this formula 
over $\Delta=\Delta_b$. Using Stokes' theorem we have
\be\label{stokey}
\int_{\Delta} R_X \diff y_1\ldots\diff y_n = -\int_{\Delta} G^{ij}_{\ \ ij} \diff y_1\ldots\diff y_n
= \sum_a \int_{\mathcal{F}_a} G^{ij}_{\ i}v^a_j\frac{1}{|v_a|}\diff\sigma-
\int_H G^{ij}_{\ i}b_j \frac{1}{|b|}  \diff\sigma
\ee
where $\diff\sigma$ denotes the measure induced on a 
hyperplane. In fact the first term on the 
right hand side of this equation is 
\be
\sum_a \frac{2}{|v_a|}\vol(\mathcal{F}_a)~.\ee%
This is easily proved using the leading behaviour of $G^{ij}$ near to 
the facets, which is universal in order that 
the metric be smooth. To see this, let 
us pick a facet, say $\mathcal{F}_1$, and take the
normal vector to be $v_1=e_1=(1,0,\ldots,0)$. Differentiating the
relation
\be\label{delta}
G^{ij}G_{jk}=\delta^i_k\ee
and setting $G=\Gcan$ we obtain
\be
(\Gcan)^{ij}_{\ \ i} \sum_a v_a^jv_a^k\frac{1}{l_a(y)} = (\Gcan)^{ij}\sum_a
v_a^iv_a^jv_a^k\frac{1}{l_a(y)^2}~.\ee
We now multiply this relation by $l_1(y)=y_1$ and take the limit
$y_1\rightarrow 0$. One obtains
\be
(\Gcan)^{1i}_{\ \ i}(y_1=0) = \lim_{y_1\rightarrow 0}\left[
  (\Gcan)^{11}\frac{1}{y_1}\right]~.\ee
Now taking the $y_1\rightarrow 0$ limit of (\ref{delta}) gives
\be
\lim_{y_1\rightarrow 0}\left[(\Gcan)^{11}\frac{1}{y_1}\right]=2\ee
and thus we obtain 
\be
(\Gcan)^{1i}_{\ \ i}(y_1=0) = 2~.\ee
The extension to general $v_1$ now follows. It should also be 
clear from this argument that setting
$G=\Gcan+\tilde{G}$ where $\tilde{G}$ is smooth on the whole of
$\mathcal{C}$ gives the same result.

On the other hand, for the second term on the right hand side of 
(\ref{stokey}) we have
\be
G^{ij}_{\ i}b_j = 2(G^{ij}G_{jk}y_k)_{,i} = 2 y_{i,i} = 2n\ee
and we thus obtain
\be
\int_{\Delta} R_X \diff y_1\ldots\diff y_n = \sum_a \frac{2}{|v_a|}\vol(\mathcal{F}_a) - 
\frac{2n}{|b|}\vol(H)~.\ee
However, we may now use the fact that \cite{lasserre}
\be
\vol(\Delta)= \frac{1}{2n|b|}\vol(H)~.\label{lassformula}\ee
This generalises the usual formula for the area of a triangle 
to higher dimensional polytopes. We give a proof of this in the next 
section.
Together with the formulae (\ref{vol}), (\ref{faces}) we thus obtain
\be\label{inty}
\int_{X_1} R_X\diff y_1\ldots\diff y_n = 
(2\pi)^n\int_{\Delta} R_X \diff y_1\ldots\diff y_n= \frac{2\pi}{(n-1)}\sum_a \vol(\Sigma_a) 
- 2n\vol(Y)~.\ee
Note that for \emph{compact} toric K\"ahler manifolds the last term is
absent and, using another result from \cite{lasserre}, one easily reproduces 
the formula in \cite{abreu2}. 
For our non--compact case of interest, we see that 
the integrated Ricci scalar of $X$ is independent of $g$. Indeed, the 
right hand side of (\ref{inty}) is manifestly 
only a function of the Reeb vector $b$. 

We may now set $R_X=0$ for 
a Ricci--flat K\"ahler metric and we thus prove the relation
\be
\pi\sum_a \vol(\Sigma_a) = n(n-1)\vol(Y)~.
\label{form}\ee
Note that in the case of regular Sasaki--Einstein manifolds this formula
in fact follows from a topological argument.

We conclude this section by deriving a relation valid for an arbitrary
polytope in $\mathbb{R}^n$. The proof is again a simple application of 
Stokes' theorem. Consider the following form of Stokes' theorem:
\bea
\int_\Delta  \nabla f \; \diff y_1\dots \diff y_n & = & \int_{\de \Delta} f \; v \, \diff \sigma
\eea
where $v$ is the outward--pointing normal vector to the boundary. 
Taking $f$ to be the constant function, and using (\ref{lassformula}), 
gives immediately
\bea
\sum_a \frac{1}{|v_a|}\mathrm{vol}({\cal F}_a) \, v_a & = & 2n\,  \mathrm{vol}(\Delta) \; b 
\label{cool}
\eea
where recall that the $v_a$ are inward pointing, and $b$ is outward pointing. 
As a first application of this result, consider the special case of a toric 
Gorenstein singularity, for which we can take the inward primitive 
normals to the facets to be of the form $v_a=(1,w_a)$. The first component 
of equation (\ref{cool}) then implies
\bea
\pi \sum_a \mathrm{vol}(\Sigma_a) & = & b_1\, (n-1) \mathrm{vol}(Y)  ~, 
\label{cool1}
\eea
where we have used (\ref{vol}) and (\ref{faces}) to pass from volumes of the 
polytope to $Y$.
Comparing this with (\ref{form}) we find that for Sasaki--Einstein 
metrics the component of the Reeb vector along the Calabi--Yau plane  must be
\bea 
b_1  =  n~. 
\eea
Notice that the same result was obtained by studying regularity of 
the Monge--Amp\`ere equation (\ref{MA}) on $\mathcal{C}$. A third derivation 
will be offered in the next section. Also note that this 
proves that the canonical metric $\Gcan_{ij}$ is never Ricci--flat 
for $d>n$, since $\bcan_1=d$. In the case $d=n$ the metric on 
$X$ is an orbifold of the flat metric on $\C^n$.


\section{A variational principle for the Reeb vector}

In this section we derive a variational principle that determines the 
Reeb vector of a Sasaki--Einstein metric in terms of the 
toric data of a fixed toric 
Gorenstein singularity. The Reeb vector is the unique 
critical point of a function
\be
Z:\mathcal{C}^*\rightarrow\R\ee
which is closely related to the volume of the polytope $\Delta$. 
Existence and uniqueness of this local minimum is proven using a simple 
convexity argument. 
We examine the extremal function in 
detail in the case $n=3$ and determine the Reeb vector in  a number of 
examples. In particular we correctly reproduce the Reeb vector and 
volumes for the explicit family of metrics $Y^{p,q}$ and also examine 
the case of the suspended pinch point and 
the complex cone over the second del Pezzo surface. 
In the latter case no Sasaki--Einstein metric is known, or even known to 
exist. Nevertheless the dual field theories are known for all these 
singularities and the corresponding volumes can be computed
in field theory using $a$--maximisation.  For the second del Pezzo surface
this computation was performed in  \cite{BBC}, which corrected previous 
results in the literature.
We find agreement with the computation obtained by extremising $Z$.

\subsection*{The extremal function}

We begin with the Einstein--Hilbert action for a 
metric $h$ on $Y$. This is given by a functional
\be
S:\mathrm{Met}(Y)\rightarrow \R\ee
which explicitly is
\be
S[h] = \int_Y \left(R_Y +2(n-1)(3-2n)\right)\diff\mu_Y~.\ee
Here $\diff\mu_Y$ is the usual measure on $Y$ constructed from the metric 
$h$ and $R_Y=R_Y(h)$ is the Ricci scalar of $h$. 
The Euler--Lagrange equation for 
this action gives the Einstein equation
\be
\mathrm{Ric}_Y(h) = (2n-2)h~.\ee
This is equivalent to the metric cone
\be
\diff s^2 (X) = \diff r^2 + r^2h\ee
being Ricci--flat. 

We would like to interpret $S$ as a functional on the space of 
\emph{Sasakian} metrics on $Y$, and use the integral formula for the 
Ricci scalar of $X$ derived in the previous section. The 
relationship between the 
Ricci scalar of the metric $h$ on $Y$ and the Ricci scalar of the 
cone $X$ over $Y$ is straightforward to derive:
\be
R_X = \frac{1}{r^2}\left[R_Y + (2n-1)-(2n-1)^2\right]~.\ee
Integrating this over $X_1$ gives
\be
\int_{X_1} R_X = \frac{1}{2n-2}\int_Y \left(R_Y + [(2n-1)-(2n-1)^2]\right)
\diff\mu\ee
and hence for a Sasakian metric $h$ we compute
\be
S[h] = 2(n-1)\left[\frac{2\pi}{n-1}\sum_a\vol(\Sigma_a)-2n\vol(Y)\right] 
+4(n-1)\vol(Y)\ee
giving
\be
S = S[b] = 4\pi\sum_a \vol(\Sigma_a)-4(n-1)^2\vol(Y)~.\ee
Remarkably we see that the action depends only on $b$. Thus we may 
interpret $S$ as a function
\be
S:\mathcal{C}^*\rightarrow\R~.\ee
Moreover, 
Sasaki--Einstein metrics are critical points of this function. Thus 
we simply impose
\be
\frac{\de}{\de b_i} S = 0\ee
which is a set of $n$ algebraic equations for $b$ in terms of only 
the toric data {\it i.e.} the normal vectors $v_a$. 

We may write the function $S$ more usefully as a function on the polytope 
$\Delta$:
\be
Z[b]\equiv\frac{1}{4(n-1)(2\pi)^n}\ S[b] = \sum_a\frac{1}{|v_a|}\vol(\mathcal{F}_a) 
-2n(n-1)\vol(\Delta)~.\ee
Using (\ref{cool1}) we can write this as
\bea\label{Zfun}
Z[b] & = & \big(b_1-(n-1)\big)\, 2n ~\mathrm{vol}(\Delta_b)~.
\eea
Notice that $Z[b]$ is then manifestly positive 
(negative) for $b_1>n-1$ ($b_1<n-1$).

It is interesting to take the derivative of $S$ along the Euler vector 
on the dual cone $\mathcal{C}^*$:
\be
b_i\frac{\de}{\de b_i}S = -2(n-1)^2\int_{X_1} R_X~.\label{eulerS}\ee
Thus we see that scalar flatness implies this component of the 
variational problem. Using (\ref{Zfun}) and the fact that $\vol(\Delta_b)$ 
is homogeneous 
degree $-n$ in $b$ we have
\be
b_i\frac{\de}{\de b_i}Z = -2n(n-1)(b_1-n)~\vol(\Delta)\ee
and this in turn implies that $b_1=n$ for a critical point. Thus 
all critical points of $Z$ lie on this plane in $\mathcal{C}^*$. 
Recall that this 
was also a necessary condition for a solution to the Monge--Amp\`ere 
equation to correspond to a smooth metric on $Y$.

\subsection*{Existence and uniqueness of an extremum}

We have shown that $b_1=n$ for all critical points of $Z$, and thus 
we may introduce a reduced function
\be
\tilde{Z}=Z\mid_{b_1=n} = 2n~\vol(\Delta)\mid_{b_1=n}~.\ee
We must now set the variation of this to zero with respect to 
the remaining variables $b_2,\ldots,b_n$. 

There is a general formula for the volume of a convex polytope, and 
in principle one can carry out this extremisation explicitly. However, 
even in dimension $n=3$ the formula for $\vol(\Delta)$ can be 
quite unwieldy. We examine this general formula 
in more detail in the next subsection. In the current subsection we would 
instead like to prove that there is always a critical point of 
$Z$ in $\mathcal{C}^*$, and moreover this critical point is unique and is a 
global minimum of $\tilde{Z}$. The critical point is therefore also 
the unique local minumum of $Z$ -- the global minimum is of course $-\infty$.
The strategy is to show that $\vol(\Delta)$ is 
a strictly convex function on $\mathcal{C}^*_0$, and then use 
standard convexity arguments to argue for a unique critical point.

Let us first assume that $\vol(\Delta)$ is a strictly convex function 
of $b$ on $\mathcal{C}^*_0$. It is simple to see that $\vol(\Delta)$ 
tends to $+\infty$ everywhere on $\partial\mathcal{C}^*$. 
Geometrically this is the limit where the characteristic hyperplane 
$H$ no longer intersects the polyhedral cone $\mathcal{C}$ to form 
a finite polytope. Also note that $\vol(\Delta)$ is bounded below by zero 
and is continuous. Hence there must be some minimum of $\tilde{Z}$ in 
the interior of the finite polytope in $\mathcal{C}^*$ defined by $b_1=n$. 
Moreover since $\vol(\Delta)$ 
is strictly convex there is a unique such critical point which is also a
global minimum of $\tilde{Z}$, and we are done.

It remains then to prove that $\vol(\Delta)$ is strictly convex 
on $\mathcal{C}^*_0$. Our proof of this is remarkably simple. 
Let us write $\Delta=\mathcal{C}\cap\{2(b,y)<1\}$, and set 
$V(b)\equiv\vol(\Delta)$. Then
\be
V = \int_{\Delta} \diff y_1\ldots\diff y_n = 
\int_{\mathcal{C}} \theta(1-2(b,y))\diff y_1\ldots\diff y_n\ee
where we have introduced the Heaviside step function 
$\theta(1-2(b,y))$. Differentiating this with respect to $b$ gives
\be
\frac{\de V}{\de b_i} = -\int_H y_i \frac{1}{|b|}\diff \sigma\ee
where recall that the characteristic hyperplane $H=\mathcal{C}\cap
\{2(b,y)=1\}$ and $\diff\sigma$ is the usual measure on the hyperplane 
$H\subset\R^n$. Here we've simply used the fact that the derivative 
of the step function is a delta function. As a check on this formula, 
one can contract with $b_i$ to obtain
\be
b_i\frac{\de V}{\de b_i} = -\frac{1}{2|b|} \vol(H)~.\ee
However by Euler's theorem the left hand side is 
simply $-nV$, and hence we have proven the relation (\ref{lassformula}) 
that we used earlier. 

We may now appeal to another result from reference \cite{lasserre}, 
which again is straightforward to prove. Since $y_i$ is homogeneous degree 
1 we have $(y_j\de/\de y_j)y_i=y_i$ and thus we compute
\be
(n+1)\int_{\Delta}y_i \ \diff y_1\ldots\diff y_n = \int_{\Delta} \frac{\de}{\de y_j}(y_jy_i) \ \diff y_1\ldots\diff y_n
= \frac{1}{2|b|}\int_H y_i\diff\sigma\ee
where in the last step we have used Stokes' Theorem and the fact that 
on $\partial\mathcal{C}$ we have $(v_a,y)=0$. Thus
\be
\frac{\de V}{\de b_i} = -2(n+1)\int_{\Delta} y_i \ \diff y_1\ldots\diff y_n~.\ee
Introducing a Heaviside function again and differentiating we thus obtain\footnote{It is straightforward to check this formula by brute force in dimension 
$n=2$.}
\be
\frac{\de^2 V}{\de b_i \de b_j} = \frac{2(n+1)}{|b|}\int_H y_i y_j\diff\sigma~.\ee
The integrand is now positive semi--definite, hence 
the Hessian of $V$ is positive definite, and so $V$ is strictly convex 
on $\mathcal{C}^*_0$.

\subsection*{The extremal function in $n=3$ and examples}

The case of most physical interest is when the toric Calabi--Yau cone
has complex dimension $n=3$, and the corresponding Sasaki--Einstein manifold 
$Y$ has real dimension five. Here we can give a simple formula 
for $Z[b]$ and the volumes 
in terms of $b$ and the toric data -- namely the primitive 
normals $v_a=(1,w_a)$ that define the polyhedral cone $\mathcal{C}$.

Denote by $v_1,\dots,v_d$ the primitive normals, ordered in such a way that the  
corresponding facets are adjacent to each other, with $v_{d+1}\equiv v_1$.  
The volume of the $a$th facet is then given by 
\bea
\frac{1}{|v_a|}\mathrm{vol}({\cal F}_a) & = & \frac{1}{8}\,\frac{(v_{a-1},v_a,v_{a+1})}{(b,v_{a-1},v_a)(b,v_a,v_{a+1})}
\eea
where $(v,w,z)$ is the determinant of the $3\times 3$ matrix whose rows 
(or columns) are $v,w$ and $z$, respectively.
The volume of the polytope can for instance 
be obtained from the first component of 
(\ref{cool})
\be
\vol(\Delta_b) = \frac{1}{6b_1}\sum_a \frac{1}{|v_a|}\vol(\mathcal{F}_a)~.\ee
Clearly this is homogeneous degree $-3$ in $b$. 
The volumes of the submanifolds $\Sigma_a$ 
and the volume of $Y$ are then determined
using the formulae given earlier. Explicitly we have
\bea
\mathrm{vol}(\Sigma_a) & = & 2\pi^2\, 
\frac{(v_{a-1},v_a,v_{a+1})}{(b,v_{a-1},v_a)(b,v_a,v_{a+1})} 
\label{volfacet3}\\
\mathrm{vol}(Y) & = & \frac{\pi^3}{b_1}\, 
\sum_a \frac{(v_{a-1},v_a,v_{a+1})}{(b,v_{a-1},v_a)(b,v_a,v_{a+1})} ~.
\label{voltot3}
\eea

\subsubsection*{The conifold}

\begin{figure}[!th]
\vspace{5mm}
\begin{center}
\epsfig{file=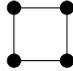,width=0.9cm,height=0.9cm}\\
\end{center}
\caption{Toric diagram for the conifold.} \label{con} \vspace{5mm}
\end{figure}

Let us start with the simplest and most familiar example of a toric 
non--orbifold 
singularity: the conifold. This is the Calabi--Yau cone over the homogeneous 
Sasaki--Einstein manifold $T^{1,1}$. The corresponding toric diagram 
is also well--known. A derivation of this starting from the conifold 
metric was presented in the Appendix of reference \cite{toricpaper}.
The inward pointing normals to the polyhedral cone in $\mathbb{R}^3$ 
may be taken to be
\bea
v_1= [1,1,1]~,\quad v_2 =[1,0,1]~,\quad v_3=[1,0,0]~,\quad v_4=[1,1,0]~.
\eea
Projecting these onto the $e_1=1$ plane one obtains the toric diagram in figure
\ref{con}.

Notice that we have listed the normal vectors in the order of the facets of 
the polyhedral cone. The corresponding 3--submanifolds $\Sigma_a$ are four copies 
of $S^3$. The extremal function is computed to be
\bea
Z[x,y,t] & = & \frac{(x-2)x}{8yt(x-t)(x-y)}
\eea
where here, and in the following examples, we set $b=(x,y,t)$. 
After imposing $x=3$ the remaining equations are then 
easily solved, and it turns out that 
there is a unique solution on $\R^3$. The extremising Reeb vector is
\bea
b_\mathrm{min} & = & \left(3,\frac{3}{2},\frac{3}{2}\right)~. 
\eea
One now easily computes
\bea
\mathrm{vol}(\Sigma_a)\,= \,\frac{8}{9} \pi^2, \qquad \frac{\pi}{6}\cdot 4 \cdot \frac{8}{9} \pi^2\,=\, \frac{16}{27}\pi^3 \,=\, \mathrm{vol}(T^{1,1})~.
\eea
These results are in fact well--known in the physics literature.

\subsubsection*{The $Y^{p,q}$ toric singularities}

\begin{figure}[!th]
\vspace{5mm}
\begin{center}
\epsfig{file=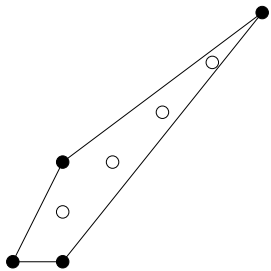,width=3.5cm,height=3.5cm}\\
\end{center}
\caption{Toric diagram for $Y^{5,3}$.} \label{Y52}
\end{figure}

The $Y^{p,q}$ toric singularities were determined in reference 
\cite{toricpaper}
by explicitly constructing the moment map for the $\T^3$ action
on the  $Y^{p,q}$ manifolds. The metrics on $Y^{p,q}$ were constructed 
in  references 
\cite{paper1,paper2}. The inward pointing normals to the four--faceted
polyhedral cone may be taken to be
\bea
v_1= [1,0,0]~,\quad v_2 =[1,p-q-1,p-q]~,\quad v_3=[1,p,p]~,\quad v_4=[1,1,0]~.
\label{ypqvbasis}
\eea
This corresponds to the basis of $\T^3$ in which the toric diagrams were
originally presented in reference
\cite{toricpaper}. Note that again we have listed the normals in the 
order of the facets of the polyhedral cone. 
%
In figure \ref{Y52} we display,  as an example, the case of $Y^{5,3}$.

We compute the following function
\be
Z[x,y,t] = 
\frac{(x-2)p(p(p-q)x+q(p-q)y+q(2-p+q)t)}{8t(px-py+(p-1)t)((p-q)y+(1-p+q)t)(px+qy-(q+1)t)}~.
\ee
Extremising this function is best left to Mathematica. Imposing $x=3$, 
the remaining equations have four solutions on $\R^3$. However, only one 
lies within the dual cone $\mathcal{C}^*$, as must be the case from 
our earlier general analysis of the function $Z$.
The final result is the following Reeb vector
\bea
b_{\mathrm{min}} & = & \left(3,\frac{1}{2}(3p-3q+\ell^{-1}),\frac{1}{2}(3p-3q+\ell^{-1})\right)
\eea
where 
\bea
\ell^{-1} &= &\frac{1}{q}(3q^2-2p^2+p\sqrt{4p^2-3q^2})~.
\eea
This is precisely the Reeb vector of the 
$Y^{p,q}$ metrics \cite{paper2,toricpaper}. One then easily reproduces the 
total volume 
\be
\mathrm{vol}(Y^{p,q}) = \frac{q^2[2p+(4p^2-3q^2)^{1/2}]}{3p^2[3q^2-2p^2+p(4p^2-3q^2)^{1/2}]}
     \pi^3\ee
and the volume of the supersymmetric 
submanifolds corresponding to the four facets 
\cite{toricpaper,quiverpaper,HEK}, respectively.

\subsubsection*{The suspended pinch point}

The suspended pinch point (SPP) is a toric 
Gorenstein singularity where the five inward pointing normals 
to $\mathcal{C}$ may be taken to be
\be
v_0 = [1,0,0]~,\quad v_1=[1,-1,0]~,\quad v_2=[1,0,1]~,\quad v_3=[1,1,1]~,\quad v_4=[1,1,0]~.\ee
Here we have also included the blow--up mode $v_0$. 

\begin{figure}[!th]
\vspace{5mm}
\begin{center}
\epsfig{file=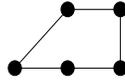,width=1.6cm,height=1cm}\\
\end{center}
\caption{Toric diagram for the SPP.} \label{SPP}
\end{figure}

%
Introducing the gauge--invariant monomials
\be
u=Z_1Z_4, \quad v=Z_2Z_3, \quad w=Z_1^2Z_2, \quad z=Z_3Z_4^2\ee
we see that an equivalent algebraic description of the singularity 
is given by the hypersurface 
\be
wz = u^2v\ee
in $\C^4$. The boundary of this conical singularity is in fact singular. 
One can see this from the normal vectors as follows. 
Clearly $<v_1,v_4>_{\mathbb{R}}\cap\Z^3$ is the sublattice $\Z^2\subset\Z^3$ 
spanned by $e_1$ and $e_2$. However, $<v_1,v_4>_{\Z}$ does not 
generate all of $\Z^2$ -- for example, one cannot generate the 
vector $(1,0,0)$. Thus the polyhedral cone is not good, in the sense 
of reference \cite{L}, and hence the boundary $Y_{SPP}$ must be singular. 
Indeed, the two vectors $v_1,v_4$ define an edge of the cone $\mathcal{C}$, 
and this edge does not satisfy the condition (\ref{good}).
In fact from the gauged linear sigma model it is easy to see 
\cite{MP} that
$Y_{SPP}$ is the cube root of the canonical circle bundle over the 
orbifold $\mathbb{C}P^1_{[1,2]}\times\mathbb{C}P^1_{[1,2]}$ where 
$\mathbb{C}P^1_{[1,2]}$ is a weighted projective space -- this 
is the symplectic quotient $\C^2//U(1)$ where the $U(1)$ has 
charges $(1,2)$.

The function $Z$ is given by   
\bea
Z[x,y,t] & = & \frac{(x-2)(2x-t)}{8t(t-x)(t-x-y)(x-y)} ~.
\label{ZSPP}
\eea
This attains its local minimum at
\bea
b_\mathrm{min} & = & \left(3,\frac{1}{2}(3-\sqrt{3}),3-\sqrt{3}\right)~.
\eea
The volume of the corresponding Sasaki--Einstein metric\footnote{This 
metric has recently been 
obtained in \cite{CLPP,MS2,CLPP2} as a member of an infinite
family of toric Sasaki--Einstein metrics generalising $Y^{p,q}$. 
The volume indeed agrees with the value presented here.} 
 is then given by
\bea\label{junior}
\mathrm{vol}(Y_{SPP}) & = & \frac{2}{9} \sqrt{3}\, \pi^3~.
\eea
We also compute the following volumes: 
\bea
\mathrm{vol}(\Sigma_1) = \mathrm{vol}(\Sigma_4)  = \frac{2}{3}\pi^2~,\qquad 
\mathrm{vol}(\Sigma_2) = \mathrm{vol}(\Sigma_3)  = \frac{2}{3}(-1+\sqrt{3})\pi^2~.
\eea
These results may be compared with the dual field theory calculations.
The gauge theory for the SPP  was obtained in reference \cite{MP}
and it is straightforward to perfom $a$--maximisation for this 
theory. Without entering into the details, we obtain the following
function to maximise:
\bea
\frac{32}{9}a(x,y,z,t) &=& 3 +
(x-1)^3+(y-1)^3+(z-1)^3+(t-1)^3\nn\\
&+&(x+y-1)^3+(1-x-y-z)^3+(1-x-y-t)^3~.
\label{amaxSPP}
\eea
Evaluating $a$ at its local 
maximum gives\footnote{We suppress factors of $N$.}
\be
a(Y_{SPP})  =  \frac{3}{8}\sqrt{3}~.\ee
\label{aSPP}
Using the AdS/CFT formula
\be \label{adscft} 
a_Y = \frac{\pi^3}{4\cdot\mathrm{vol}(Y)}\ee
we therefore find perfect agreement with the geometrical result 
(\ref{junior}). It is quite remarkable 
that extremisation of the
function $Z$ in (\ref{ZSPP}) and $a$ in (\ref{amaxSPP}) are two completely
equivalent problems.

\subsubsection*{The complex cone over $dP_2$}

In the following we determine the Reeb vector for the Sasaki--Einstein 
metric\footnote{Assuming that it exists.}
on the boundary of the complex cone over the second del Pezzo surface, 
$dP_2$. Recall that a del Pezzo surface $dP_k$ is the blow--up 
of $\mathbb{C}P^2$ at $k$ generic points. 
Recall also that the 
first two del Pezzo surfaces do not admit K\"ahler--Einstein metrics 
\cite{tian, tianyau}. This fact follows straightforwardly 
from Matsushima's theorem \cite{matsushima}. 
Thus the boundaries of 
the complex
cones over $dP_1$ and $dP_2$ must be non--regular Sasaki--Einstein manifolds. 
In fact 
in \cite{toricpaper} it was shown 
that $Y^{2,1}$ is an irregular metric for the case of $dP_1$, while the metric 
for the case of $dP_2$ remains
unknown. Nevertheless, using our extremisation method 
one can determine the volume for this metric.

\begin{figure}[!th]
\vspace{5mm}
\begin{center}
\epsfig{file=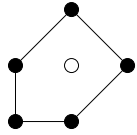,width=1.7cm,height=1.7cm}\\
\end{center}
\caption{Toric diagram for the complex cone over $dP_2$.} \label{dP2} 
\end{figure}
\noindent
The five inward pointing normals may be taken to be
\be
v_1 = [1,0,0]~,\quad v_2=[1,0,1]~,\quad v_3=[1,1,2]~,\quad v_4=[1,2,1]~,
\quad v_5=[1,1,0]~.\ee
%
The extremal function in this basis is
\bea
Z[x,y,t] & = & \frac{(x-2)(-t^2+2t(x+y)+(3x-y)(x+y))}{8yt(t-x-y)(t+x-y)(t-3x+y)}~.
\eea
The extremum that lies inside the dual cone is computed to be
\bea
b_\mathrm{min} & = & \left(3,\frac{9}{16}(-1+\sqrt{33}), \frac{9}{16}(-1+\sqrt{33})\right)~.
\eea

We may now compute the volume of the corresponding Sasaki--Einstein metric:
\bea
\mathrm{vol}(Y_{dP_2}) & = & \frac{1}{486} (59+11\sqrt{33})\, \pi^3~.
\eea
This agrees with the value 
for this volume predicted by the authors of \cite{BBC} using the purely field 
theoretic technique of $a$--maximisation together with the AdS/CFT 
formula (\ref{adscft}).
We also compute the following volumes 
\bea
\mathrm{vol}(\Sigma_1) = \frac{1}{81}(17+\sqrt{33})\pi^2~,\quad
\mathrm{vol}(\Sigma_2) = \mathrm{vol}(\Sigma_5)=\frac{1}{27}(1+\sqrt{33})\pi^2~,\nn\\
\mathrm{vol}(\Sigma_3) = \mathrm{vol}(\Sigma_4)=\frac{2}{81}(9+\sqrt{33})\pi^2~.
\eea
It is then straightforward to match these with the R--charges of 
fields computed in reference \cite{BBC}.

\section{Discussion}

In this paper we have shown that, for a given toric 
Calabi--Yau cone, the problem of determining the Reeb vector
for the Sasaki--Einstein metric on the base of the cone 
is decoupled from that 
of finding the metric itself. The Reeb vector is determined by 
finding the unique critical point to the function
\be
Z:\mathcal{C}^*\rightarrow \R~.\ee
It is then easy to see that this information is sufficient 
to compute the volume of the Sasaki--Einstein manifold, as well as the volumes
of toric submanifolds which are complex divisors
in the corresponding Calabi--Yau cone. For illustrative 
purposes, we have solved
explicitly the extremal problem in a number of examples in complex 
dimension $n=3$. One 
would also like to prove uniqueness and existence of 
a solution $g\in\mathcal{H}(1)$ of the Monge--Amp\`ere equation (\ref{MA2}) 
to complete the analysis of toric Sasaki--Einstein manifolds, 
but we leave this for future work.

In the case of $n=3$ it is interesting to compare the geometrical 
results of this paper with $a$--maximisation in superconformal 
gauge theories in four dimensions. In order to do this, let 
us reformulate the extremal problem in the following way. 
A generic Reeb vector may be written
\bea
b &=& b_0 + \sum_{i=2}^n s_i b_i\eea
where $b_0=ne_1$, $b_i=e_i$, $i=2,\ldots,n$, and 
$s_i\in\R$. The vector $b_0$ is such that the $(n,0)$--form 
$\Omega$ of the Ricci--flat metric 
has charge $n$ under the corresponding Killing vector field, 
whereas the $b_i$ leave $\Omega$ invariant. Indeed, recall that all 
critical points of $Z$ necessarily lie on the plane $(b,e_1)=n$. 
The Reeb vector for the 
Sasaki--Einstein metric is then the unique global minimum of the 
reduced function $\tilde{Z}$, now regarded as a function of the 
parameters $s_i$. Moreover at the critical point, $\tilde{Z}$ and 
$Z$ are just the volume of the Sasaki--Einstein metric, up 
to a dimension--dependent factor.

Recall now that, starting 
from a toric Calabi--Yau singularity in complex dimension 
three, one can construct a four--dimensional supersymmetric quiver gauge 
theory arising from a stack of $N$ D3--branes 
placed at the singularity, 
which is expected to flow at low energies to a non--trivial 
superconformal fixed point. The Higgs branch of this 
gauge theory is essentially the toric Calabi--Yau singularity. 
$a$--maximisation allows one to 
fix uniquely the exact R--symmetry of this theory at the 
infra--red fixed point. This may be formulated as follows. 
One first fixes a fiducial R--symmetry $R_0$ which satisfies 
the constraints imposed by anomaly cancellation. This R--charge 
is then allowed to mix with the set of global abelian non--R 
symmetries of the theory -- by definition the supercharges are 
invariant under these symmetries. Thus the trial R--symmetry may be 
written as \cite{IW}
\bea
R &=& R_0 + \sum_I s_I F_I\eea
where $F_I$ generate the group of abelian symmetries, and $s_I\in\R$. 
One can now define a function 
$a$ which is a sum over a cubic function of the
R--charges of fields in the theory, and is
thus a function of the $s_I$. 
The exact R--symmetry of the theory at its conformal fixed point is 
uniquely determined by (locally) maximising this function $a$ over the space 
of $s_I$ \cite{IW}.
Moreover, the value of $a$ at the critical point is precisely the $a$--central charge 
of the gauge theory, which is inversely proportional to the volume 
of the dual Sasaki--Einstein manifold via the AdS/CFT formula (\ref{adscft}).

Now, the AdS/CFT correspondence states that the subgroup of the 
isometry group of the Sasaki--Einstein manifold that commutes with 
the Reeb vector is precisely the set of flavour symmetries of the 
dual gauge theory. Recall that we showed that the Reeb vector 
cannot mix with the non--abelian part of the isometry group. 
In complex dimension $n=3$, this is the geometrical realisation of the 
field theory statement that the R--symmetry does not mix with non--abelian 
factors of the global symmetry group of the gauge theory \cite{IW}.
Therefore
the minimisation of $Z$ may always be performed over a space that is
at most \emph{two--dimensional}. Moreover, the $b_i$, $i=1,2$ precisely 
generate the $U(1)\times U(1)$ isometry under which 
the $(3,0)$--form is uncharged 
and are thus dual to flavour symmetries $F_I$ in the gauge theory. 
In contrast, note that 
$a$--maximisation is generally performed over a larger parameter space,
which includes the baryonic symmetries. However, the results here suggest 
that, for toric quiver gauge theories, it is possible to perform 
$a$--maximisation over a two--parameter space of flavour symmetries.

Notice that the problem of determining $b_{\mathrm{min}}$ is reduced to finding
the roots of polynomials whose degree generically increases with $d$, 
implying that the volumes, and hence also 
charges, of the dual theories are in general 
\emph{algebraic numbers}. 
Although all theories considered in examples so far have been found to admit
quadratic irrational charges, it is easy to see that more general algebraic
numbers are expected as a result of maximising a cubic function of more than one
variable.
The precise relation between $Z$ and $a$ for a given toric singularity remains
rather mysterious. It is clear that obtaining a 1--1 map between these 
two functions, and the details of the two extremal problems, would improve
our understanding of some aspects of these superconformal field theories. 
Tackling  this problem will require a better understanding
of how the geometric data is translated into field theory quantities. One
can anticipate that 
such quantities must be \emph{invariants} with respect to the possible choices 
of toric phase or other field theory dualities.

Finally, we would like to emphasise that our
results are valid in any dimension, while $a$--maximisation holds
only for duals of five--dimensional 
Sasaki--Einstein geometries. However, the AdS/CFT correspondence
predicts that $AdS_4\times Y_7$ geometries in
M--theory, with $Y_7$ a Sasaki--Einstein seven--manifold, are dual to 
\emph{three--dimensional} ${\cal N}=2$ 
superconformal field theories. The results
of this paper therefore suggest that there should exist some analogue of 
$a$--maximisation 
for three--dimensional theories as well. 
If true, the details of the argument should
differ substantially from those used in reference 
\cite{IW} -- in three dimensions there exist no 
anomalies to match. It will be very interesting to pursue this 
direction and explore the possibility that a field theoretic dual of
$Z$--minimisation can be formulated for superconformal field
theories in three dimensions.

\subsection*{Acknowledgments}
\noindent J. F. S. and D. M. would like to thank A. Hanany and 
B. Wecht for interesting discussions. We also thank 
T. Wiseman for assistance with Mathematica, and J.--X. Fu for pointing 
out a typographical error in an earlier version of this paper. D. M. would also like to thank 
the Physics Department of Harvard University for enjoyable hospitality while
this work was being completed.
J. F. S. is supported by NSF grants DMS--0244464, DMS--0074329 and DMS--9803347. S.--T. Y. is supported in part by NSF grants DMS--0306600 and DMS--0074329.

\end{document}